\documentclass[preprintnumbers,amsmath,amssymb]{revtex4-2}
\usepackage{color}
\usepackage{graphicx}
\usepackage{epsfig}
\usepackage{epstopdf,hyperref}
\usepackage{bm}
\newtheorem{corollary}{Corollary}[section]

\newtheorem{theorem}{Theorem}[section]
\newtheorem{proposition}{Proposition}[section]

\DeclareMathOperator{\sech}{sech}

\begin{document}
\title{Dissipative localised structures for the complex Discrete Ginzburg-Landau equation}
	
\author{Dirk Hennig}
	\author{Nikos I. Karachalios}
	\affiliation{Department of Mathematics, University of Thessaly, Lamia GR35100, Greece}
\author{Jes\'{u}s Cuevas-Maraver}
\affiliation{Grupo de F\'{i}sica No Lineal, Departamento de F\'{i}sica Aplicada I,
	Universidad de Sevilla. Escuela Polit\'{e}cnica Superior, C/ Virgen de \'{A}frica, 7, 41011-Sevilla, Spain \\
	Instituto de Matem\'{a}ticas de la Universidad de Sevilla (IMUS). Edificio Celestino Mutis. Avda. Reina Mercedes s/n, 41012-Sevilla, Spain}
\date{\today}

\begin{abstract}
The discrete complex Ginzburg-Landau  equation is a fundamental model for the dynamics of nonlinear lattices incorporating competitive dissipation and energy gain effects. Such mechanisms are of particular importance for the study of survival/destruction of localised structures in many physical situations. In this work, we prove that  in the discrete complex Ginzburg-Landau equation dissipative solitonic waveforms persist for significant times by introducing a dynamical transitivity argument. This argument is based on a combination of the notions of  ``inviscid limits'' and of the ``continuous dependence of solutions on their initial data'', between the dissipative system and its Hamiltonian counterparts.  Thereby, it establishes closeness of the solutions of the Ginzburg-Landau lattice to those of the conservative ideals described by the Discrete Nonlinear Schr\"odinger  and Ablowitz-Ladik  lattices. Such a closeness holds when the initial conditions of the systems are chosen to be sufficiently small in the suitable metrics and for small values of the dissipation or gain strengths.
Our numerical findings are  in excellent agreement with the analytical predictions for the dynamics of the dissipative bright, dark or even Peregrine-type solitonic waveforms.

\end{abstract}

\maketitle
\section{Introduction}


%

\noindent
The complex discrete Ginzburg-Landau equation (DGL), in its standard local version
\begin{equation}
\frac{d u_n}{dt}=u_n+(1+i\alpha)
(u_{n+1}-2u_n+u_{n-1})-(1+i\beta)\,|u_{n}|^2u_n,\;\;u_n \in {\mathbb{C}}\,\,\,n\in {\mathbb{Z}},\alpha,\beta\in\mathbb{R},\label{eq:GLE}
\end{equation}
is one of the fundamental discrete dissipative nonlinear lattices appearing in numerous physical contexts, such as  the dynamics of coupled waveguides, lasers and of low-dimensional fluid dynamical systems \cite{UGL1}-\cite{AGL2}. An important feature of the DGL lattice is that is connected with the physically significant Discrete Nonlinear Schr\"odinger equation (DNLS) \cite{HennigTsironis}-\cite{DNLS},
\begin{equation}
i\frac{d\phi_n}{dt}=-\alpha(\phi_{n+1}+\phi_{n-1}-2\phi_n)+\beta |\phi_n|^2\phi_n,\label{eq:discrete}
\end{equation}
as $\alpha,\beta \rightarrow \infty$.  Dividing (\ref{eq:GLE}) and (\ref{eq:discrete}) by
$\alpha$, scaling the time as $\tau=\alpha t$ and denoting $\epsilon=1/\alpha$ and $\kappa=\beta/\alpha$, we get that the DGL \eqref{eq:GLE} and the DNLS \eqref{eq:discrete} can be rewritten respectively, in the form
\begin{equation}
i\frac{d u_n}{d\tau}=i\epsilon u_n+(i\epsilon-1)
(u_{n+1}-2u_n+u_{n-1})-(i\epsilon-\kappa)\,|u_{n}|^2 u_n,\,\,\,n\in {\mathbb{Z}},\label{eq:scaledGLE}
\end{equation}
and
\begin{equation}
i\frac{d\phi_n}{d\tau}=-(\phi_{n+1}+\phi_{n-1}-2\phi_n)+\kappa |\phi_n|^2 \phi_n.\label{eq:scaleddiscrete}
\end{equation}
Hence, equation (\ref{eq:scaleddiscrete}) is obtained from (\ref{eq:scaledGLE}) in the limit $\epsilon \rightarrow 0^+$, and consequently, (\ref{eq:scaledGLE}) can be viewed as a dissipative extension of (\ref{eq:scaleddiscrete}).
Notice that if $\alpha,\beta \rightarrow \infty$, the value of $\kappa$ is determined by the rate at which $\alpha$ and $\beta$ diverge, and that for large but finite values of $\alpha$, $\beta$ being of the same order,
one has that $\kappa \sim {\cal{O}}(1)$. The sign of the parameter $\kappa$ renders the DNLS system as focusing ($\kappa<0$) or defocusing ($\kappa>0$).

The conservative DNLS equation \eqref{eq:scaleddiscrete}, being non-integrable, has attracted tremendous attention as a universal model for the dynamics of localised structures in discrete media. The existence of discrete breathers and discrete solitonic structures for the DNLS \eqref{eq:scaleddiscrete} has been established by wide blend of computational  and analytical methods, ranging from the anticontinuous limit approaches to homoclinic, asymptotic and variational methods; see the representative works \cite{MackayAubry}-\cite{PanosImaRev} and references therein. It is crucial to remark that a major problem concerning the existence of travelling solitons of DNLS is that a localised state can be pinned due to the Peierls-Nabarro barrier \cite{PN3}. This is not the case for the Ablowitz-Ladik (AL) lattice
\begin{equation}
i\frac{d\psi_n}{dt}=-(\psi_{n+1}+\psi_{n-1}-2\psi_n)+\kappa(\psi_{n+1}+\psi_{n-1})|\psi_{n}|^2,\,\,\,n\in {\mathbb{Z}},\;\;\label{eq:AL}
\end{equation}
which is integrable (by the Inverse Scattering Transform Method \cite{Ablowitz}), admits a great variety of analytical solitary solutions \cite{AL,AL2,akhm_AL,akhm_AL2}, and possesses continuous translation symmetry which allow them to travel along the lattice.

Since the existence of discrete localised modes concerns the conservative ideals \eqref{eq:scaleddiscrete} and \eqref{eq:AL}, an important question, both from the mathematical and the physical applications viewpoint, is the following: {\em to what extent may discrete solitonic structures  persist or survive in the presence of linear or nonlinear gain/loss effects, which are relevant to a more realistic description of physical set-ups where such effects cannot be neglected?}

Key works on the numerical identification and the construction of analytical or approximative soliton solutions for DGL systems which incorporate a mixture of cubic and other power-law type and saturable nonlinearities, and/or non-local terms similar to the non-local nonlinearity of the AL-system are \cite{UGL2}-\cite{UGL6}. Apart from their physical significance and relevance, the presence of the non-local terms in the DGL systems seems to be important in identifying/constructing localised modes bifurcating from solutions arising in the integrable AL-limit. Extensions of such approaches to 2D systems are provided in \cite{UGL3,UGL4,BUGL4}.

In the present paper, we aim at answering the above question by investigating the potential persistence and evolution properties of localised structures for the DGL equation \eqref{eq:scaledGLE}, by taking advantage of our recent results on the congruence of the dynamics of solutions of nonlinear lattices involving local and non-local nonlinearities \cite{DNJ2022},\cite{DN2021}.
In particular, via a {\em transitivity argument}, we establish a closeness result between the solutions of the dissipative DGL equation \eqref{eq:scaledGLE} and the solutions of the AL lattice \eqref{eq:AL} in the limit $\epsilon\rightarrow 0$ and when the initial conditions are sufficiently close in the $l^2$ or $l^{\infty}$-metric and sufficiently small in the relevant norms.

Let as analyse this  transitivity argument further.  The first step is to prove that for any  $\epsilon>0$, solutions of the DGL \eqref{eq:scaledGLE} are close to the solutions of the DNLS \eqref{eq:scaleddiscrete} in the following sense of ``a continuous dependence on their initial data'': If their initial conditions satisfy $||u_0-\phi_0||_{l^2}\leq K_0\varepsilon$, for any $0<\varepsilon<1$, $\epsilon\leq\varepsilon$,  for some constant $K_0>0$, then the corresponding solutions satisfy $||u(t)-\phi(t)||\leq C\varepsilon$, for all $t\in [0,T_f]$, for arbitrary $0<T_f<\infty$  and the constant is of the form  $C=C(u_0,\phi_0,\kappa,T_f)$. In the limit $\varepsilon\rightarrow 0$, we revisit the ``inviscid" limit of the DGL \eqref{eq:scaledGLE}, that is the DNLS \eqref{eq:scaleddiscrete} (see also \cite{Zhao} in the case of the dissipative DNLS). In the second step, by implementing the closeness results of \cite[Theorem 1.1, pg. 349]{DNJ2022}, we are able to use for our purposes the ``continuous dependence'' result, this time between the solutions of the DNLS \eqref{eq:scaleddiscrete} and the AL-lattice \eqref{eq:AL}: Assuming that their initial conditions satisfy $|| \phi_0-\psi_0||_{l^2}\le C_0\, \varepsilon$, for any $\varepsilon>0$, then we may prove an estimate of the form $|| \phi(t)-\psi(t)||_{l^2}\le \tilde{C} \varepsilon$ where again
$\tilde{C}=\tilde{C}(\phi_0, \psi_0,\kappa,T_f)$.  Consequently, combining the above two results, establishes when $\epsilon\rightarrow 0$ in the DCGL equation \eqref{eq:scaledGLE}, the convergence of its solutions to those of the AL-lattice \eqref{eq:AL}, when the distance of their initial data $||u^0-\psi^0||_{l^2}\rightarrow 0$, as $\varepsilon\rightarrow 0$, through analytical estimates in the same ``continuous dependence of their initial data'' sense as described above.

The impact of this result is that the analytical solitary solutions of the Hamiltonian AL-lattice \eqref{eq:AL} should persist for finite intervals which can be sufficiently large (pending on the closeness and smallness of the initial data of the systems), in the dissipative DGL lattice  \eqref{eq:scaledGLE}, for small values of $\epsilon>0$. {\em In particular, the corresponding dissipative localised structures should share for the time of their survival, a functional form and characteristics which should be close to the analytical solutions of the AL lattice.} These should include bright and dark solitons, even discrete rational solutions such as the discrete Peregrine soliton. The results are valid for all the distinct regimes of dissipation or gain effects present in the DGL, which affect the maximal interval of existence of solutions.

At this point, it is important to remark that in the DGL \eqref{eq:scaledGLE} the choice of a positive value for $\epsilon$ is crucial. When $\epsilon>0$, it ensures global existence of the solutions for all initial data, as the case $\epsilon>0$ corresponds to the case of linear gain and nonlinear loss  manifested by the terms $\epsilon u_n$  and $\epsilon |u_n|^2u_n$, respectively. To study the dynamics of the localised structures in different regimes of gain/loss we consider the variant of \eqref{eq:scaledGLE}
\begin{equation}
i\frac{d u_n}{dt}=i\epsilon_1 u_n+(i\epsilon_2-1)
(u_{n+1}-2u_n+u_{n-1})-(i\epsilon_3-\kappa)\,|u_{n}|^2 u_n,\,\,\,|\epsilon_i|=\epsilon>0,\;\; i=1,2,3.\label{eq:scaledGLEN}
\end{equation}
While the closeness results described above are valid for the DGL \eqref{eq:scaledGLEN} in the case where $|\epsilon_i|=\epsilon\rightarrow 0$, the time interval  $[0,T_f]$ for the study of the dynamics should be finite with $T_f<T_{\mathrm{max}}$ where $[0,T_{\mathrm{max}})$ is the generic maximal interval of existence of solutions of Eq. \eqref{eq:scaledGLEN}. This restriction is due to the fact that in the regimes of linear and nonlinear gain or linear loss and nonlinear gain, the solutions may collapse (blow-up) in finite time $T_{\mathrm{max}}$.  This is in vast contrast to the Hamiltonian DNLS and AL lattices for which solutions exist for all times. In light of such  a behaviour, we numerically test the theoretical results for the DGL system \eqref{eq:scaledGLEN} in the distinct regimes of gain and loss. The findings of the numerical experiments are in excellent agreement with the analytical predictions for the closeness of solutions and the global dynamics of the dissipative localised structures in the aforementioned regimes.

The presentation of the paper is as follows: In Section II we prove the analytical closeness/convergence result of solutions between the DCGL \eqref{eq:scaledGLEN} and the AL lattice \eqref{eq:AL}. Section III is devoted to the numerical studies. In the first part of Section III, we prove analytical results on the global asymptotic behaviour of solutions for finite lattice approximations, for all linear and nonlinear gain/loss regimes. These analytical results are particularly useful, in explaining the observed dynamics for the dissipative localised structures illustrated by the findings of the direct numerical simulations. The latter are presented in the second part of Section III.  We would like to remark, {\em that such type of results highlights the importance of the issue of the potential global asymptotic stability or instability of solutions:} Exact analytical solutions (or other approximate solutions), if constructed for the DGL equation (see for example \cite{EX1, EXA1}, \cite[pg. 208]{EX2}) and slightly perturbed, they should obey the global dynamics as analysed herein, and exhibit in long term, the instability manifested by either decay, collapse or convergence to other asymptotic states described by the global attractor of the system. In Section IV we summarise the main results and discuss potential future studies.
%
%
%
%
%
%
\section{Inviscid limit and closeness theorem}
\paragraph{Preliminaries: local and global existence of solutions.} The functional setting and properties of linear and nonlinear operators involved in the nonlinear lattices considered in the present paper is described in \cite{,DNJ2022,DN2021} and references therein.  In particular, for the DGL system \eqref{eq:scaledGLEN}, we may prove the following local existence result of solutions in the sequence spaces
\begin{equation}
l^p=\left\{ u=(u_n)_{n \in {\mathbb{Z}}}\,\in {\mathbb{C}}\,\,\,\vert \, || u||_{l^p}=\left(\sum_{n \in {\mathbb{Z}}}|u_n|^p\right)^{1/p}\right\},\;\; 1\leq p\leq \infty.
\end{equation}
\begin{proposition}
	\label{thloc}	
	Let $|\epsilon_i|=\epsilon>0$, $\kappa\in\mathbb{R}$ and  the initial condition  $u(0)=u^0\in l^2$ be arbitrary. There exists some $T_{\mathrm{max}}(u^0)>0$, such that the DGL system \eqref{eq:scaledGLEN} has a unique solution $u\in C^1([0,T_f],l^2)$  for all $0<T_f<T_{\mathrm{max}}$. In addition, the following alternatives hold: Either $T_{\mathrm{max}}=\infty$ (global existence) or $T_{\mathrm{max}}<\infty$ and $\lim_{t\uparrow T_{\mathrm{max}}}||u(t)||_{\ell^2}=\infty$ (collapse or blow-up in finite time). Furthermore, the solution $u$ depends continuously on the initial condition $u^0\in l^2$, with respect to the norm of $C([0,T_f],l^2)$.
\end{proposition}
Let us recall, that the proof of Proposition \ref{thloc} makes use of the continuous embeddings
\begin{equation}
l^r\subset l^s,\,\,\,|| w||_{l^s}\le ||
w||_{l^r},\,\,\,1 \le r\le s \le \infty,\label{eq:embeddings}
\end{equation}
which will also be used in the sequel, for the derivation of various estimates.

It is crucial to remark that a vast contrast between the DGL system \eqref{eq:scaledGLEN}, the Hamiltonian DNLS \eqref{eq:scaleddiscrete} and the AL lattice \eqref{eq:AL}, is that solutions of the DNLS exists globally in time, unconditionally with respect to the size of the initial data and the sign of the parameter $\kappa$.  This is due to the conservation of the following quantities for the DNLS system \eqref{eq:scaleddiscrete}, namely the power $P$ and Hamiltonian $H$
\begin{eqnarray*}
	P&=&\sum_{n\in\mathbb{Z}}|\phi_n|^2,\\	
	H&=&\sum_{n\in\mathbb{Z}}\left(|\phi_{n+1}-\phi_n|^2+\frac{\kappa}{2}|\phi_n|^4\right),\;\;
\end{eqnarray*}	
which for the AL-system \eqref{eq:AL} are the modified power $P_{\tiny{\mathrm{AL}}}$ and the Hamiltonian $ H_{\tiny{\mathrm{AL}}}$
\begin{eqnarray*}
P_{\tiny{\mathrm{AL}}}&=&\sum_{n\in\mathbb{Z}}\ln\left(1-\kappa|\psi_n|^2\right),\\	
 H_{\tiny{\mathrm{AL}}}&=&\sum_{n}
 \overline{\psi}_n(\psi_{n+1}+\psi_{n-1}).
\end{eqnarray*}	
The sign of parameter $\kappa$ determines whether the nonlinearity is focusing ($\kappa<0$) or defocusing ($\kappa>0$). Instead of Proposition \ref{thloc}, for the DNLS and AL lattices we have the following one.
\begin{proposition}
	\label{thlocDNLSAL}	\begin{enumerate}
		\item
	Let  $\kappa\in\mathbb{R}$ in the DNLS equation \eqref{eq:scaleddiscrete} (focusing or defocusing) and $\kappa<0$ in the AL equation \eqref{eq:AL} (focusing), and the initial conditions  $\phi(0)=\phi^0,\;\;\psi(0)=\psi^0\in l^2$  be arbitrary. Then, their unique solutions exist globally in time, i.e., $\phi,\;\psi\in  C^1([0,\infty),l^2)$. In particular the corresponding solutions satisfy
\begin{eqnarray}
\label{DNbound1}
||\phi(t)||^2_{l^2}&=&||\phi^0||_{l^2}^2,\;\;\mbox{for all $t\in [0,\infty)$},\\
\label{ALbound2}
||\psi(t)||^2_{l^2}&\leq& C_{\kappa}||\psi^0||_{l^2}^2,\;\;\mbox{for all $t\in [0,\infty)$},
\end{eqnarray}
for some constant $C_{\kappa}>0$ which depends on the parameter $\kappa$.
\item Consider the defocusing AL-equation  \eqref{eq:AL} with $\kappa>0$. Assume that the initial condition $\psi(0)=\psi^0$ satisfies the following assumptions:
\begin{eqnarray}
\label{ged}
||\psi^0||_{\infty}<\frac{1}{\kappa}\;\;\mbox{and}\;\;P_{\mathrm{AL}}(0)<\infty.
\end{eqnarray}
Then, its unique solution exist globally in time, i.e., $\psi\in  C^1([0,\infty),l^2)$ and satisfies the estimate
\begin{eqnarray}
\label{ALbound3}
||\psi(t)||^2_{l^2}&\leq& C'_{\kappa}||\psi^0||_{l^2}^2,\;\;\mbox{for all $t\in [0,\infty)$},
\end{eqnarray}
for some constant $C'_{\kappa}>0$ which also depends on the parameter $\kappa$.
\end{enumerate}
\end{proposition}

The proof of Proposition \ref{thlocDNLSAL}, concerning in particular the AL lattice follows by implementing the arguments of \cite[Lemma 2.1 and Proposition 3.1]{DNJ2022}, and we omit the details.
\paragraph{Proofs of the closeness results.} We are ready to proceed to the statements and proofs of the main analytical results of the paper. First, we prove the closeness of the solutions between the DGL equation \eqref{eq:scaledGLEN} and the DNLS equation \eqref{eq:scaleddiscrete} in the sense of a ``continuous dependence on their initial data''.
This is the first step for the transitivity argument described in the introduction.

\begin{theorem}
Let $0<|\epsilon_i|=\epsilon<1$ be arbitrary and $[0,T_{\mathrm{max}})$ be the maximal interval of existence for the solutions of the equation \eqref{eq:scaledGLEN}. Consider any finite $T_f$ such that $0<T_f<T_{\mathrm{max}}$. For any $0<\varepsilon<1$ such that $\epsilon\leq\varepsilon$, there exist a positive constant  $C(\kappa,T_f,||u^0||_{l^2},||\phi^0||_{l^2})$,
such that for initial data $u(0)=u^0$ and $\phi(0)=\phi^0$
satisfying
\begin{equation}
 ||u^0-\phi^0||_{l^2}\le K_0\varepsilon,\label{eq:delta0}
\end{equation}
the corresponding solutions of the  DGL equation \eqref{eq:scaledGLEN} and the (scaled) DNLS equation (\ref{eq:scaleddiscrete})  satisfy  for any $t\in [0,T_f]$,
\begin{eqnarray}
|| u(t)-\phi(t)||_{l^2}&\le& C \varepsilon.
 \label{eq:scaledbound}
\end{eqnarray}
\label{Theorem:scaledcloseness}
\end{theorem}
{\bf Proof:}
  We  shall use the distance variable $\Delta_n=u_n-\phi_n$ between the solutions.  Subtracting  (\ref{eq:scaleddiscrete}) from  \eqref{eq:scaledGLEN} we see  that it satisfies the equation
 \begin{eqnarray}
 \label{dif1}
 \frac{d\Delta_n}{dt}=i\epsilon_1u_n+(i\epsilon_2-2)\Delta_n-i\epsilon_3|u_n|^2u_n+\kappa\left[|u_n|^2u_n-|\phi_n|^2\phi_n\right].
 \end{eqnarray}
Multiplying \eqref{dif1} by $\overline{\Delta_n}$, summing over $\mathbb{Z}$ and keeping imaginary parts, we derive:
\begin{eqnarray}
\frac{d || \Delta||_{l^2}^2}{dt}=2|| \Delta||_{l^2}\frac{d || \Delta||_{l^2}}{dt}=&&\epsilon_1\mathrm{Re}\sum_{n\in\mathbb{Z}}u_n\overline{\Delta_n}
+\epsilon_2\sum_{n \in \mathbb{Z}}|\Delta_{n+1}-\Delta_n|^2\nonumber\\
&-&\epsilon_3\mathrm{Re}\sum_{n\in\mathbb{Z}}|u_n|^2u_n\overline{\Delta_n}+\kappa\mathrm{Im}\sum_{n\in\mathbb{Z}}\left[|u_n|^2u_n-|\phi_n|^2\phi_n\right]\overline{\Delta_n}.
\label{eq:closeness}
 \end{eqnarray}
Due to Proposition \ref{thloc}, there exists a constant $M=M(\kappa,T_f,||u^0||_{l^2})$, such that the solutions of the DGL \eqref{eq:scaledGLEN} satisfy
\begin{eqnarray}
\label{dif2}
||u||_{l^2}\leq M,\;\;\forall t\in [0,T_f],
\end{eqnarray}
while the solutions of the DNLS (\ref{eq:scaleddiscrete}) satisfy the conservation \eqref{DNbound1}. With these bounds, Cauchy-Schwarz inequality and the inclusions relation \eqref{eq:embeddings}, we may estimate each term on the right-hand side of \eqref{eq:closeness}. For the first and third term, by using the bound \eqref{dif2}, we get the estimates
\begin{eqnarray}
\label{dif3}
\left|\epsilon_1\mathrm{Re}\sum_{n\in\mathbb{Z}}u_n\overline{\Delta_n}\right|&\leq& \epsilon ||u||_{l^2}||\Delta||_{l^2}\leq \epsilon M||\Delta||_{l^2},\\
\label{dif4}
\left|\epsilon_3\mathrm{Re}\sum_{n\in\mathbb{Z}}|u_n|^2u_n\overline{\Delta_n}\right|&\leq& \epsilon\left(\sum_{n\in\mathbb{Z}}|u_n|^6\right)^{\frac{1}{2}}\left(\sum_{n\in\mathbb{Z}}|\Delta_n|^2\right)^{\frac{1}{2}}=\epsilon||u||_{l^6}^3||\Delta||_{l^2}\leq \epsilon ||u||_{l^2}^3||\Delta||_{l^2}\leq \epsilon M^3||\Delta||_{l^2}.
\end{eqnarray}
For the fourth term, we use the bound \eqref{DNbound1} and the inequality
\begin{eqnarray*}
	\left||u_n|^2u_n-|\phi_n|^2\phi_n\right|\leq |\phi_n|^2|u_n-\phi_n|+|u_n|\left(|\phi_n|+|u_n|\right)|u_n-\phi_n|,
\end{eqnarray*}	
to derive the estimate
\begin{eqnarray}
	\label{dif5}
	\left|\kappa\mathrm{Im}\sum_{n\in\mathbb{Z}}\left[|u_n|^2u_n-|\phi_n|^2\phi_n\right]\overline{\Delta_n}\right|&\leq& |\kappa|\,||\phi||_{l^{\infty}}^2||\Delta||^2_{l^2}+|\kappa|\,||u||_{l^{\infty}}\left(||\phi||_{l^{\infty}}+||u||_{l^{\infty}}\right)||\Delta||^2_{l^2}\nonumber\\
	&\leq& |\kappa|\,||\phi||^2_{l^{2}}||\Delta||^2_{l^2}+|\kappa|\,||u||_{l^{2}}\left(||\phi||_{l^{2}}+||u||_{l^{2}}\right)||\Delta||^2_{l^2}\nonumber\\
	&\leq& |\kappa|\,||\phi^0||^2_{l^{2}}||\Delta||^2_{l^2}+|\kappa|\,M\left(||\phi^0||_{l^{2}}+M\right)||\Delta||^2_{l^2}\nonumber\\
	&&=M_0||\Delta||_{l^2}^2, \;\;M_0=|\kappa|\left[||\phi^0||_{l^2}(1+M)+M^2\right].
\end{eqnarray}
We set the constants $M_1=M+M^3$, $M_1=M_1(\kappa,T_f,||u^0||_{l^2})$ and $M_2=M_0+4\epsilon$, where $M_0=M_0((\kappa,T_f,||u^0||_{l^2},||\phi^0||_{l^2})$, as it can be seen from \eqref{dif5}. Then, from \eqref{eq:closeness} and the estimates \eqref{dif3}-\eqref{dif5}, we see that $||\Delta||_{l^2}$ satisfies the linear differential inequality
 \begin{equation}
 \label{dif7}
  \frac{d || \Delta||_{l^2}}{dt}\le \epsilon M_1+M_2 || \Delta ||_{l^2}.
 \end{equation}
Integration of \eqref{dif7} (or Gronwall's inequality), and the assumption \eqref{eq:delta0} on the initial data gives:
 \begin{eqnarray}
 \label{dif8}
||\Delta(t)||_{l^2}&\leq& ||\Delta(0)||_{l^2}\exp(M_2t)+\epsilon\frac{M_1}{M_2}\left[\exp(M_2t)-1\right]\nonumber\\
&\leq&
\varepsilon K_0\exp(M_2t)+\epsilon\frac{M_1}{M_2}\left[\exp(M_2t)-1\right],\;\;\forall t\in [0, T_f].
\end{eqnarray}
Thus, since $\epsilon\leq\varepsilon$, the estimate stated in \eqref{eq:scaledbound} is valid with the constant
\begin{equation}
\label{CD}	
 C= K_0\exp(M_2T_f)+\frac{M_1}{M_2}\left[\exp(M_2T_f)-1\right],
\end{equation}
implying also the corresponding limit in \eqref{eq:scaledbound}, and the proof is completed.\ \ \ $\square$

Regarding second-order systems, we refer to \cite{Pelinovsky1} where error estimates for the
approximation of the dynamics of a diatomic infinite Fermi–Pasta–Ulam (FPU) system with light
and heavy particles by the dynamics of the monoatomic
FPU system for a small mass ratio were derived.

The exponential dependence of the constant $C$ given in \eqref{CD}, on $T_f$, is not surprising: we refer also to the time-growth estimates for the relevant distance function between the solutions of the complex Ginzburg-Landau pde and the NLS pde, when the inviscid limit of the former is considered \cite{JWU}, which can even grow exponentially \cite{OG}.

The second step of the transitivity argument is based on the following result concerning the closeness of solutions between the DNLS and the AL systems, in the sense of ``continuous dependence of initial data'', discussed above. It is proved in  \cite[Theorem 1.1, pg. 349]{DNJ2022}.

Analytical estimates for $T_{\mathrm{max}}$ in the case of the blow-up regime, will be proved in Section III for the case of finite lattices.
\begin{theorem}
	\label{Lemma:closeness}
	Consider the DNLS  equation \eqref{eq:scaleddiscrete}
	and the AL equation \eqref{eq:AL}, and assume that their initial data satisfy the conditions of Proposition \ref{thlocDNLSAL} for global in time existence.  We further assume that for every $0<\varepsilon<1$,
	their initial conditions satisfy:
	\begin{eqnarray}
		\label{eq:distance0}
		|| \phi^0-\psi^0||_{l^2}&\le& C_0\, \varepsilon^3,\\
		\label{eq:distance01}
		|| \phi^0||_{l^2}&\le& C_{\phi^0}\, \varepsilon,\label{eq:phi0}\\
		\label{eq:Pmu0}
		P_{AL}(0)&=&\sum_{n}\ln\left(1+\kappa|\psi_n(0)|^2\right)\le \ln\left(1+(C_{AL}\,\varepsilon)^2\right),\label{eq:PAL}
	\end{eqnarray}
	for some constants $C_0, C_{\phi_0},C_{AL}>0$.	
	Then, for arbitrary finite $0<T_{\small{f}}<\infty$, there exists a constant $\tilde{C}=\tilde{C}(\phi_0,\psi_0,\kappa, C_0,T_{\small{f}})$, such that the corresponding solutions
	for every $t\in [0,T_{\small{f}}]$, satisfy the estimate 	
	\begin{equation}
		|| \phi(t)-\psi(t)||_{l^2}\le \tilde{C} \varepsilon^3.\label{eq:boundy}
	\end{equation}
\end{theorem}

One of the main applications of Theorem \ref{Lemma:closeness} is that the DNLS equation admits small-amplitude  solutions, of the order ${\cal{O}}(\varepsilon)$, that stay ${\cal{O}}(\varepsilon^3)$-close to the soliton solutions of the AL equation. We remark that the results of \cite{DNJ2022}, are valid for both the focusing and defocusing cases of the DNLS and AL lattices.

Combining Theorems \ref{Theorem:scaledcloseness} and \ref{Lemma:closeness}, we conclude with the proof of the transitivity argument, which gives the main result of the paper.

\begin{theorem}
	Let the assumptions (\ref{eq:delta0})  and
	(\ref{eq:distance0})-(\ref{eq:PAL})  of Theorems \ref{Theorem:scaledcloseness} and \ref{Lemma:closeness}, hold.
	For any $0<\epsilon\leq \varepsilon <1$ and any $T_f$  as described in Theorem\ref{Theorem:scaledcloseness}, there exist  a positive constant $K_1>0$ such that
	$||u^0-\psi^0||\leq K_1\varepsilon$, and a positive constant  $C_1=C_1(u^0, \phi^0, \psi^0, \kappa, T_f)$
	such that the associated solutions of the  DGL equation(\ref{eq:scaledGLEN}) and the AL equation (\ref{eq:AL})  satisfy  for any $t\in [0, T_f]$, the estimate
	\begin{eqnarray}
		|| u(t)-\psi(t)||_{l^2}&\le& C_1 \varepsilon.
		\label{eq:scaledboundAL}
	\end{eqnarray}
	\label{Theorem:scaledclosenessAL}
\end{theorem}
{\bf Proof:} Due to the assumptions (\ref{eq:delta0}) and (\ref{eq:distance0}), the initial conditions $u^0$ and $\psi^0$ satisfy,
\begin{eqnarray*}
	||u^0-\psi^0||_{l^2}\leq ||u^0-\phi^0||_{l^2}+||\phi^0-\psi^0||_{l^2}\leq K_0\varepsilon+C_0\varepsilon^3\leq K_1\varepsilon,	
\end{eqnarray*}	
for some $K_1>0$. Then, with the aid of the estimates (\ref{eq:scaledbound}) and (\ref{eq:boundy}), we derive
\begin{eqnarray}
	|| u(t)-\psi(t)||_{l^2}&\le&|| u(t)-\phi(t)||_{l^2}+|| \phi(t)-\psi(t)||_{l^2} \nonumber\\
	&\le& C\varepsilon + \tilde{C}\varepsilon^3\leq C_1\varepsilon,
\end{eqnarray}
where $\phi(t)$ is the solution of the DNLS equation (\ref{eq:scaleddiscrete}), and $C_1$ is some constant depending on $C$ and $\tilde{C}$,
and the proof is finished.\ \ \ $\square$


Theorem \ref{Theorem:scaledclosenessAL} establishes, when $|\epsilon_i|=\epsilon\rightarrow 0$ in the DCGL equation \eqref{eq:scaledGLEN}, its solutions convergence to those of the AL-lattice \eqref{eq:AL}, when the distance of their initial data $||u^0-\psi^0||_{l^2}\rightarrow 0$, as $\varepsilon\rightarrow 0$.  In terms of applications in the context localised structures in discrete media, Theorem \ref{Theorem:scaledclosenessAL} certifies the persistence of dissipative discrete solitons for finite time intervals when the DGL equation \eqref{eq:scaledGLEN} is the underlying evolution equation for the system. This persistence will be illustrated in the next section, by direct numerical simulations, where actually, we will investigate numerically the consequences of the following corollary.
\begin{corollary}
	\label{JQ}
	Assume that the initial conditions of the AL equation \eqref{eq:AL} and of the DGL equation \eqref{eq:scaledGLEN} satisfy $u^0=\psi^0$ and $||\psi^0||_{l^2}\leq C_{\psi^0}\varepsilon$, for any $0<\varepsilon<1$, $\epsilon\leq\varepsilon$, for some constant $C_{\psi^0}>0$. Then there exists a positive constant $C_1=C_1(u^0, \psi^0, \kappa, T_f)$, such that
	the associated solutions of the  DGL equation(\ref{eq:scaledGLEN}) and the AL equation (\ref{eq:AL})  satisfy  for any $t\in [0, T_f]$, the estimate
	\begin{eqnarray*}
		|| u(t)-\psi(t)||_{l^2}&\le& C_1 \varepsilon.
	\end{eqnarray*}
\end{corollary}
\textbf{Proof:} We formally fix the initial condition of the DNLS equation \eqref{eq:scaleddiscrete}, so that $\phi^0=\psi^0$ in order to apply the transitivity argument. Note that since $u^0=\psi^0$, the requirement $||u^0-\psi^0||_{l^2}\leq K_1\varepsilon$ is trivially satisfied for any $K_1>0$ and any $0<\varepsilon<1$. Note also that since $||\psi^0||_{l^2}\leq C_{\psi^0}\varepsilon$, then  due to the elementary inequality $\ln(1+bx)\leq bx$, for all $b>0$, $x>0$, the condition \eqref{eq:PAL} is satisfied too. Hence, the result of the
Corollary follows immediately from Theorem \ref{Theorem:scaledclosenessAL}. \ \ $\Box$

It is also important to remark, that due to the congruence results between the local DGL \eqref{eq:scaledGLEN} and its nonlocal counterpart, discussed in \cite{DN2021}, which are valid even in the absence of external forcing $g_n=0$, for all $n\in \mathbb{Z}$ (see \cite[Eq. (1.3)]{DN2021}), the nonlocal DGL exhibits the same dynamics as described by Theorem \ref{Theorem:scaledclosenessAL} and Corollary \ref{JQ}, under similar smallness conditions on its initial data.

 \section{Numerical studies}
 \begin{figure}[tbp]
 	\begin{tabular}{cc}
 		\includegraphics[width=9cm]{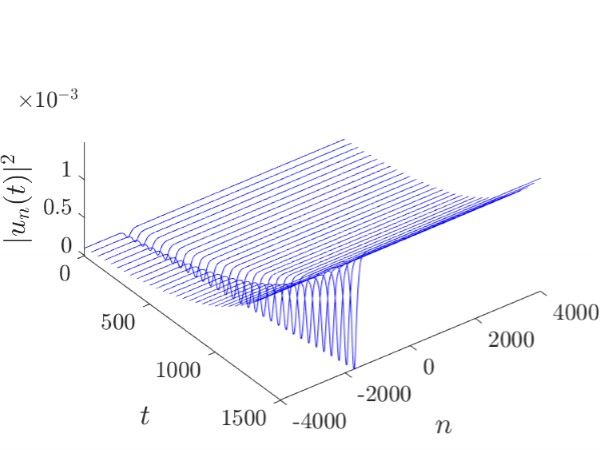} &
 		\includegraphics[width=8.5cm]{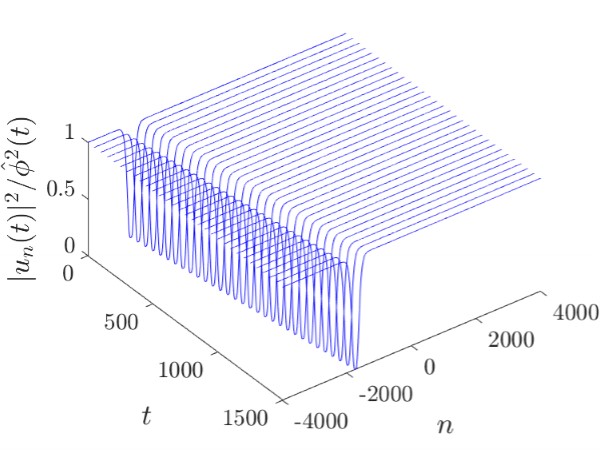} \\
 		\includegraphics[width=8.8cm]{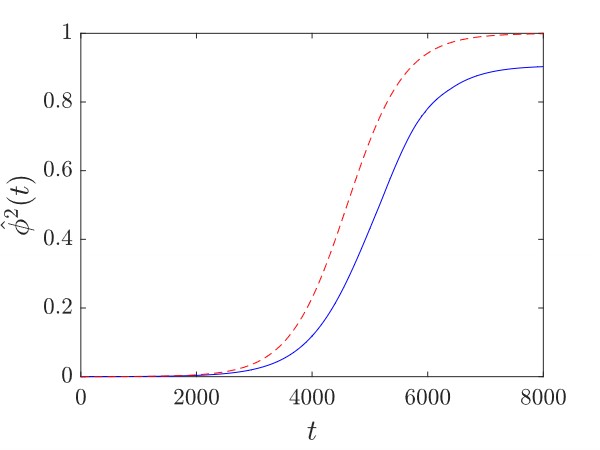} &
 		\includegraphics[width=8.8cm]{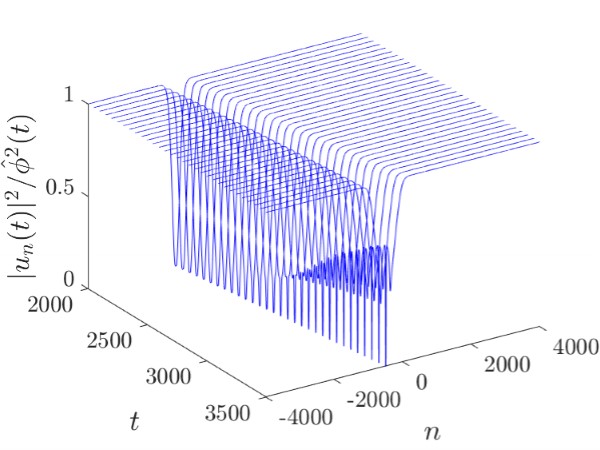} \\
 	\end{tabular}%
 	\caption{Defocusing nonlinearity ($\kappa=1>0$) for the DGL lattice \eqref{eq:scaledGLEN} with $\epsilon_2=0.001$, in the case of linear gain $\epsilon_1=0.001$ and nonlinear loss $\epsilon_3=0.001$. Spatiotemporal evolution of the initial condition $u_n(0)=\psi_n^d(0)$ defined by the the dark soliton analytical solution \eqref{eq:dark-soliton} of the AL lattice \eqref{eq:AL},  for $\beta=0.01$ and $\alpha=\pi/10$ (details in the text-see Section $\mathrm{B.a.1}$).}
 	\label{fig:evolution_dark}
 \end{figure}
\setcounter{equation}{0}
In this section, we report on the results of  numerical studies regarding the persistence and asymptotic behaviour of dissipative solitary waves in the DGL equation (\ref{eq:scaledGLEN}). Different regimes for linear/nonlinear gain or loss terms lead to  distinct dynamics for the solutions, as it will be demonstrated in the first section dealing with finite lattice approximations induced by the application of periodic or Dirichlet boundary conditions, being necessary for the implementation of numerical schemes. We note that the closeness results of Section II remain valid when the system is supplemented with the above mentioned boundary conditions. In the second section, we illustrate the numerical findings for the dynamics of bright and dark solitons and discrete rational solutions, when used as initial conditions for the DGL system \eqref{eq:scaledGLEN}.
\begin{figure}[tbp]
	\begin{tabular}{cc}
		\includegraphics[width=9cm]{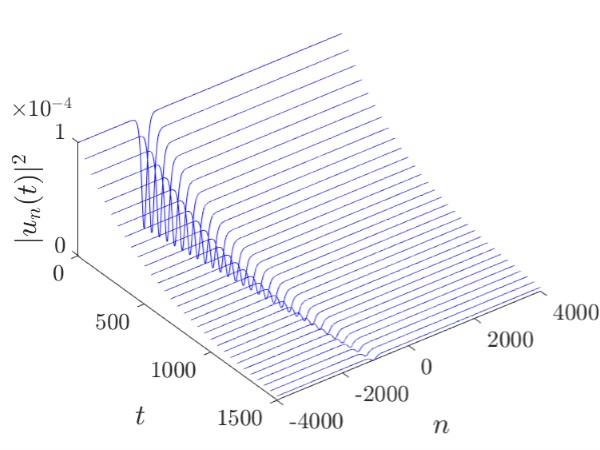} &
		\includegraphics[width=9cm]{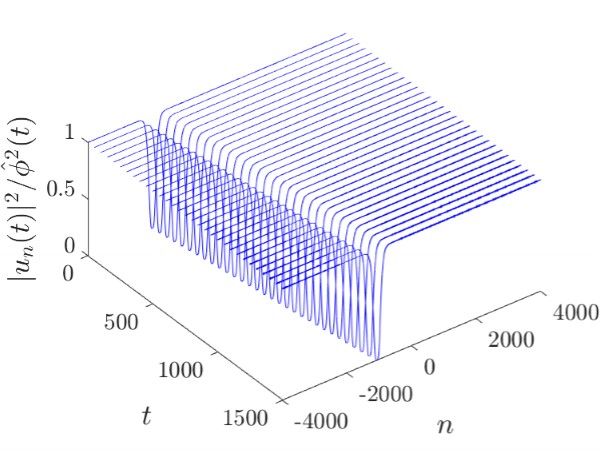}
	\end{tabular}%
	\caption{Defocusing case $\kappa>0$ for the DGL lattice \eqref{eq:scaledGLEN} with $\epsilon_2=0.001$ in the case of linear loss $\epsilon_1=-0.001$ and nonlinear loss $\epsilon_3=0.001$. Initial conditions and parameters as in Figure \ref{fig:evolution_dark} (details in the text-see Section $\mathrm{B.a.2}$).}
	\label{fig:evolution_dark1}
\end{figure}
\begin{figure}[tbp]
	\begin{tabular}{cc}
		\includegraphics[width=9cm]{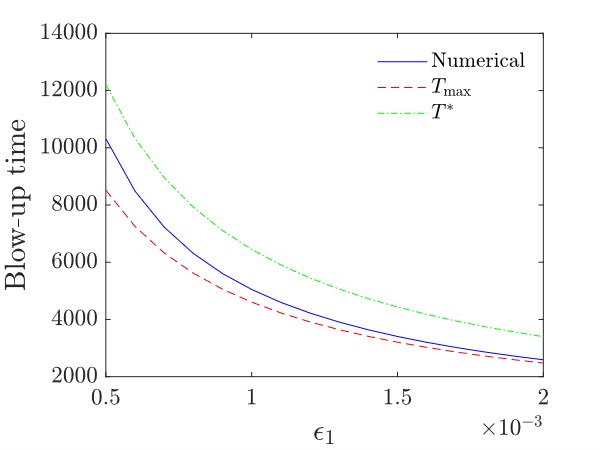} &
		\includegraphics[width=9cm]{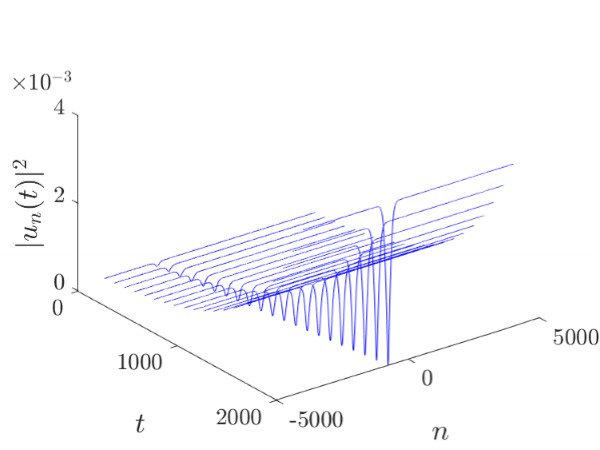}
	\end{tabular}%
	\caption{Blow-up regime: Defocusing case $\kappa>0$ for the DGL lattice \eqref{eq:scaledGLEN} with $\epsilon_2=0.001$, nonlinear gain $\epsilon_3=-0.001$ ($\tilde{\epsilon}_3=0.001$) and linear gain $\epsilon_1>0$. Left panel: Numerical blow-up time as a function of $\epsilon_1$ [continuous (blue) curve] against the analytical upper bound \eqref{enes1} [dotted-dashed (green) curve] and the analytical blow-up time for plane waves \eqref{ans8} [dashed (red) curve]. Right panel: The evolution of the dark-soliton with $\epsilon_1=0.001$ towards blow-up. Details in the text-see Section $\mathrm{B.a.3}$.}
	\label{fig:blow-up}
\end{figure}
\begin{figure}[tbp]
	\begin{tabular}{cc}
		\includegraphics[width=9cm]{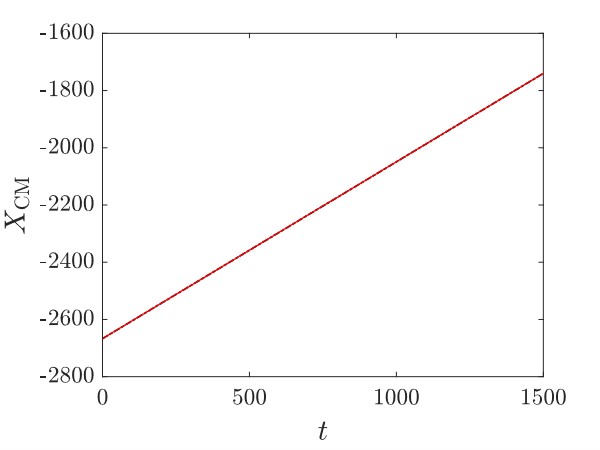} &
		\includegraphics[width=9cm]{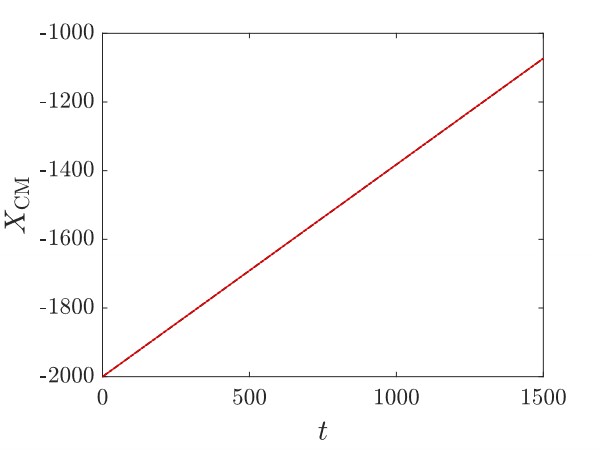} \\
	\end{tabular}%
	\caption{Evolution of the center of mass for the dissipative dark soliton in the regime of linear gain/nonlinear loss studied in Figure \ref{fig:evolution_dark} (left panel), and in the regime of linear loss/nonlinear loss studied in Figure \ref{fig:evolution_dark1} (right panel). Dashed black line corresponds to the AL-lattice analytical dark soliton and red full line holds for the evolution in the DGL equation \eqref{eq:scaledGLEN}.}
	\label{fig:CM_dark}
\end{figure}
\subsection{Set-up and comments on the asymptotic behaviour of solutions of the finite-lattice approximations}
In the numerical simulations we consider a finite lattice occupying a symmetric interval $[-L,L]$, where the position of the $N+1$ equidistantly placed oscillators is given by the discrete spatial coordinate 	$x_n=-L+nh$, $n= 0,1,2,\ldots,N$, with $h=2L/N$ being the lattice spacing. The finite lattice is supplemented by either periodic boundary conditions $u_n = u_{n+N}$ or Dirichlet boundary conditions $u_0=u_N=0$, and the system is considered in the finite dimensional spaces
\begin{eqnarray*}
{l}^p_{\mathrm{per}}:=\left\{U=(U_n)_{n\in\mathbb{Z}}\in\mathbb{R}:\quad U_n=U_{n+N},\quad
\|U\|_{l^p_{\mathrm{per}}}:=\left(h\sum_{n=0}^{N-1}|U_n|^p\right)^{\frac{1}{p}}<\infty\right\}, \quad 1\leq p\leq\infty,\\
{l}^p_{0}:=\left\{U=(U_n)_{n\in\mathbb{Z}}\in\mathbb{R}:\quad U_0=U_{N}=0,\quad
\|U\|_{l^p_{0}}:=\left(h\sum_{n=1}^{N-1}|U_n|^p\right)^{\frac{1}{p}}<\infty\right\}, \quad 1\leq p\leq\infty,
\end{eqnarray*}
respectively. The case $h=O(1)$  corresponds to the discrete regime of the system and the case $h\rightarrow 0$ approximates the continuous regime.  The norms in the case of the finite dimensional subspaces will  be denoted by $l^p$ for simplicity, and are equivalent according to the inequality
\begin{eqnarray}
\label{equi}
||U||_{\ell^q}\leq ||U||_{\ell^p}\leq N^{\frac{(q-p)}{qp}}||U||_{\ell^q},\;\;1\leq p\leq q<\infty.
\end{eqnarray}
\begin{figure}[tbp]
	\begin{tabular}{cc}
		\includegraphics[width=9cm]{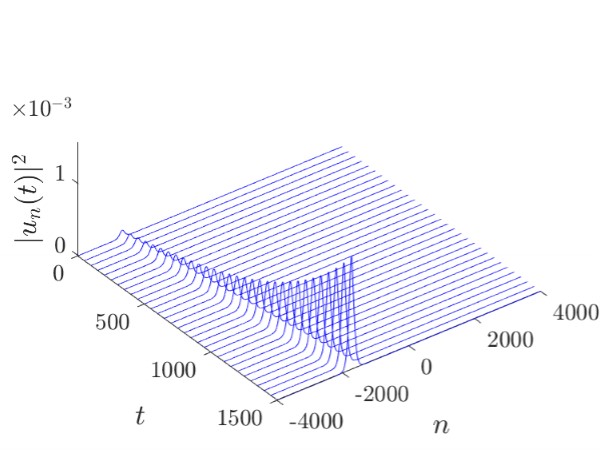} &
		\includegraphics[width=9cm]{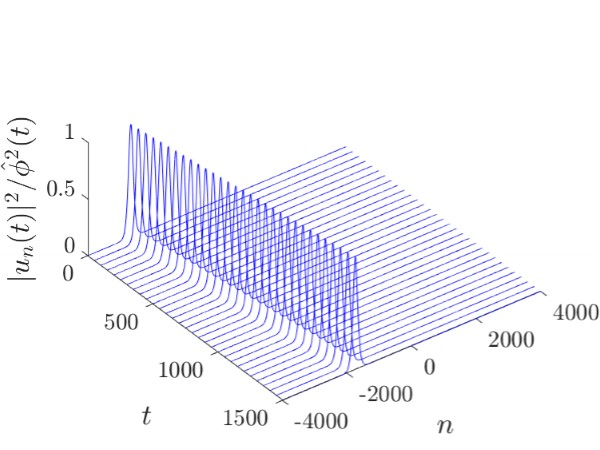}
	\end{tabular}%
	\caption{Focusing case $\kappa<0$ for the DGL lattice \eqref{eq:scaledGLEN} with $\epsilon_2=0.001$ in the case of linear gain $\epsilon_1=0.001$ and nonlinear loss $\epsilon_3=0.001$. Spatiotemporal evolution of the initial condition $u_n(0)=\psi_n^s(0)$ defined by the the bright soliton analytical solution \eqref{eq:one-soliton} of the AL lattice \eqref{eq:AL},  for $\beta=0.01$ and $\alpha=\pi/10$ (details in the text-see Section $\mathrm{B.b.1}$).}
	\label{fig:evolution_bright}
\end{figure}
\begin{figure}[tbp]
	\begin{tabular}{cc}
		\includegraphics[width=9cm]{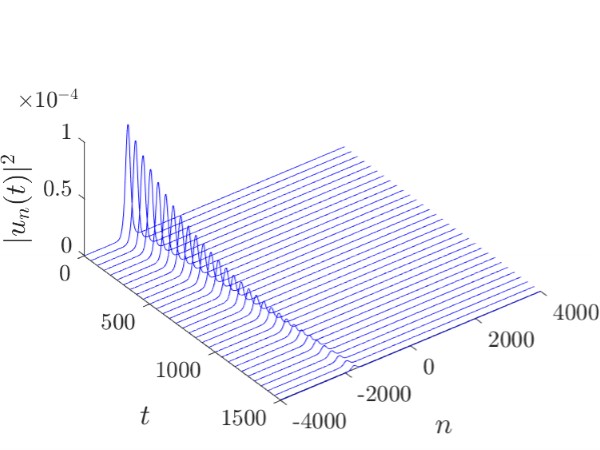} &
		\includegraphics[width=9cm]{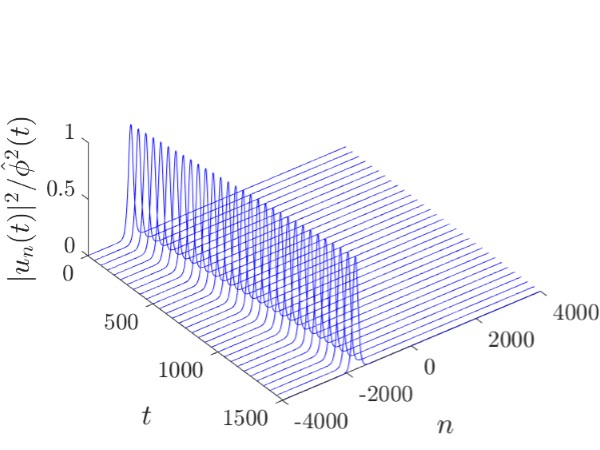}
	\end{tabular}%
	\caption{Focusing case $\kappa<0$ for the DGL lattice \eqref{eq:scaledGLEN} with $\epsilon_2=0.001$ in the case of linear loss $\epsilon_1=-0.001$ and nonlinear loss $\epsilon_3=0.001$. Initial conditions and parameters as in Figure \ref{fig:evolution_bright} (details in the text-see Section $\mathrm{B.b.2}$).}
	\label{fig:evolution_bright1}
\end{figure}
\begin{figure}[tbp]
	\begin{tabular}{cc}
		\includegraphics[width=9cm]{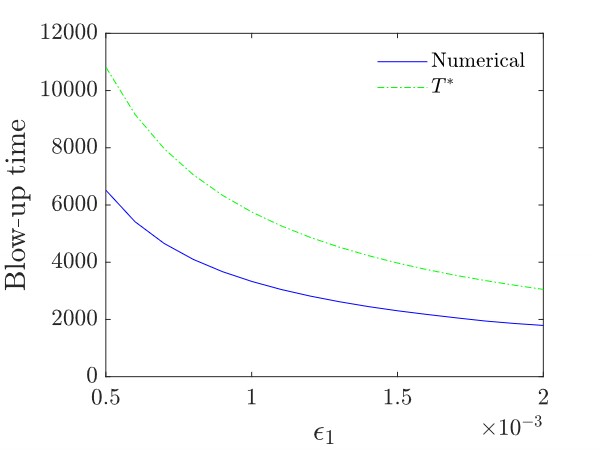} &
		\includegraphics[width=9cm]{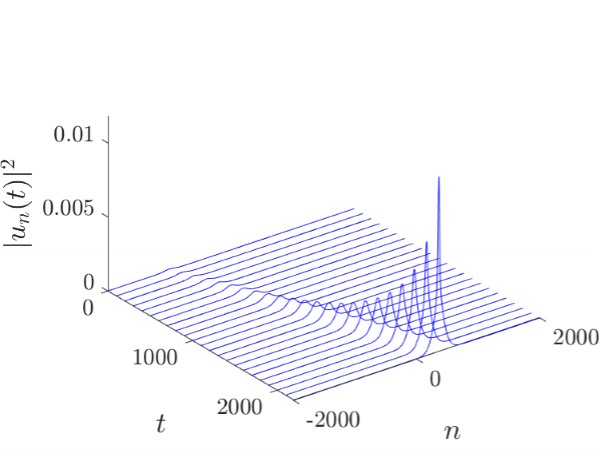}
	\end{tabular}%
	\caption{Blow-up regime: Focusing case $\kappa<0$ for the DGL lattice \eqref{eq:scaledGLEN} with $\epsilon_2=0.001$, nonlinear gain $\epsilon_3=-0.001$ ($\tilde{\epsilon}_3=0.001$) and linear gain $\epsilon_1>0$. Left panel: Numerical blow-up time as a function of $\epsilon_1$ [continuous (blue) curve] against the analytical upper bound \eqref{enes1}. Right panel: The evolution of the bright soliton with $\epsilon_1=0.001$ towards blow-up.  Details in the text-see Section $\mathrm{B.b.3}$.}
	\label{fig:blow-up2}
\end{figure}
%
%
\begin{figure}[tbp]
	\begin{tabular}{cc}
		\includegraphics[width=9cm]{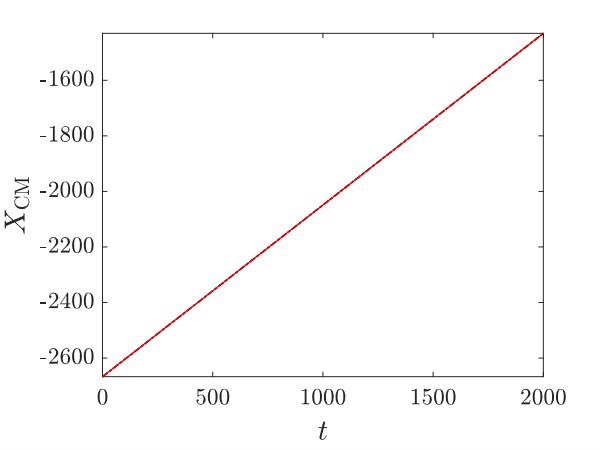} &
		\includegraphics[width=9cm]{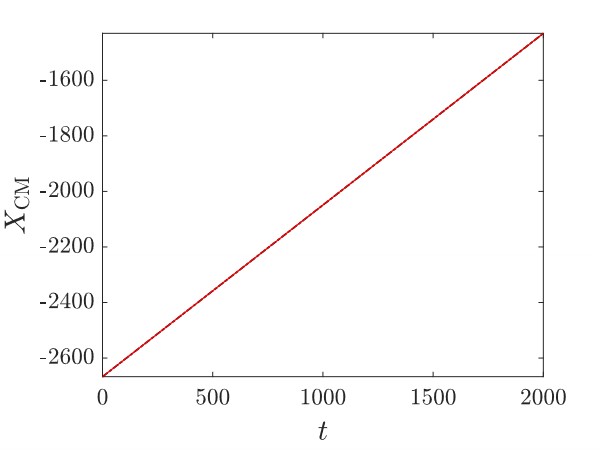} \\
	\end{tabular}%
	\caption{Evolution of the center of mass for the dissipative bright soliton in the regime of linear gain/nonlinear loss studied in Figure \ref{fig:evolution_bright} (left panel), and in the regime of linear loss/nonlinear loss studied in Figure \ref{fig:evolution_bright1} (right panel). Dashed black line corresponds to the AL-lattice analytical bright soliton and red full line holds for the evolution in the DGL equation \eqref{eq:scaledGLEN}.}
	\label{fig:CM_bright}
\end{figure}
We will also use the spatially-averaged $l^2$-norm of the solutions. The square of this norm will be denoted by
\begin{eqnarray*}
	\chi(t)=\frac{h}{N}\sum_{n=0}^{N-1}|u_n(t)|^2,\;\; \mbox{(periodic boundary conditions)},\\
	\chi(t)=\frac{h}{N}\sum_{n=1}^{N-1}|u_n(t)|^2,\;\; \mbox{(Dirichlet boundary conditions)}.
\end{eqnarray*}
Regarding  the implementation of the above boundary conditions, when the analytical soliton solutions of the AL system are inserted as initial data, the boundary conditions are strictly satisfied only asymptotically, as $L\rightarrow\infty$.
However for a sufficiently large $L$ the induced error has
negligible effects on the observed dynamics.

Prior to the presentation of the numerical results, it is important to explain  the asymptotic behaviour of the above finite lattice approximations. For brevity we choose the case $\epsilon_2>0$ which induces stronger dissipation to the system. We comment on the case $\epsilon_2\leq 0$ after the following proof.
\begin{theorem} In the DGL equation \eqref{eq:scaledGLEN}, we fix $\epsilon_2>0$ and impose on the lattice either periodic or Dirichlet boundary conditions. We distinguish between the following cases for the maximal time-interval of existence of solutions $[0, T_{\mathrm{max}})$:
\label{FAP}
\begin{enumerate}
	\item Linear gain $\epsilon_1>0$ and nonlinear loss $\epsilon_3>0$: Then the solutions exist globally in time, i.e.,  $T_{\mathrm{max}}=\infty$,  are uniformly bounded with respect to time and  the spatially averaged $l^2$-norm of the solution satisfies
	\begin{eqnarray}
	\label{est1}
\chi(t)\leq \bigg[\frac{e^{-2\epsilon_1t}}{\chi(0)}+\frac{\epsilon_3}{\epsilon_1}\left(1-e^{-2\epsilon_1t}\right)\bigg]^{-1},\;\;\;\;\;	
	\limsup_{t\rightarrow\infty}\chi(t)\leq \frac{\epsilon_1}{\epsilon_3}.
	\end{eqnarray}
\item Linear loss $\epsilon_1=-\tilde{\epsilon}_1<0$ and nonlinear loss $\epsilon_3>0$. We have $T_{\mathrm{max}}=\infty$, and the solutions of the system decay,
\begin{eqnarray}
\label{est2}
\chi(t)\leq \bigg[\frac{e^{2\tilde{\epsilon}_1t}}{\chi(0)}+\frac{\epsilon_3}{\tilde{\epsilon}_1}\left(1-e^{2\tilde{\epsilon}_1t}\right)\bigg]^{-1},\;\;\;\;\;
\lim_{t\rightarrow\infty}\chi(t)=0.
\end{eqnarray}	
	\item Linear gain or linear loss $\epsilon_1 \geq 4\epsilon_2-\tilde{\epsilon}_3||u^0||^2$ and nonlinear gain $\epsilon_3=-\tilde{\epsilon}_3<0$. For all initial data $u^0\in l^2$, the solutions of the system collapse (blow-up) in finite time $T_{\mathrm{max}}$ satisfying the estimate
\begin{eqnarray}
\label{estim3}
T_{\mathrm{max}}&\leq& \frac{1}{2(\epsilon_1-4\epsilon_2)}\ln\left[1+\frac{\epsilon_1-4\epsilon_2}{\tilde{\epsilon}_3 ||u^0||^2_{l^2}}\right],\;\;\epsilon_1>4\epsilon_2-\tilde{\epsilon}_3||u^0||^2:=\epsilon_1^{\mathrm{crit}},
\\
\label{estim3a}
T_{\mathrm{max}}&\leq& \frac{1}{\tilde{\epsilon}_3||u^0||^2_{l^2}},\;\;\;\epsilon_1=4\epsilon_2.
\end{eqnarray}	
\end{enumerate}	
\end{theorem}
{\bf Proof:} We consider only the case of periodic boundary conditions since the case of Dirichlet boundary conditions can be proved exactly in the same manner.\\
1.  Multiplying \eqref{eq:scaledGLEN} by $\overline{u}_n$ in the $l^2$-inner product and taking imaginary parts, we get that $\chi (t)$ satisfies
\begin{eqnarray}
\label{feq1}
\dot{\chi}+\frac{2\epsilon_2}{N}h\sum_{n=0}^{N-1}|u_{n+1}-u_{n}|^2=2\epsilon_1\chi-\frac{2\epsilon_3}{N}h\sum_{n=0}^{N-1}|u_n|^4.
\end{eqnarray}
Applying the inequality \eqref{equi} for $q=4$ and $p=2$ implies
\begin{eqnarray}
\label{feq2}
||u||_{l^2}^4\leq N||u||^4_{l^4},
\end{eqnarray}
which can be used to estimate the last term of \eqref{feq1}, and we arrive at the differential inequality for $\chi$,
\begin{eqnarray}
\label{feq3}
\dot{\chi}&\leq& 2\epsilon_1\chi-2\epsilon_3\left(\frac{h}{N}\sum_{n=0}^{N-1}|u_n|^2\right)^2\nonumber\\
&&=2\epsilon_1\chi-2\epsilon_3\chi^2,
\end{eqnarray}
which is of Bernoulli type, and can be integrated as follows: Using the change of variables $u=\chi^{-1}$ in \eqref{feq3}, we see that $u$ satisfies
\begin{eqnarray}
\label{feq4}
\dot{u}\geq -2\epsilon_1 u+2\epsilon_3,\label{eq:dotu}
\end{eqnarray}
and with the aid of the integrating factor $e^{2\epsilon_1t}$, (\ref{eq:dotu}) can be solved, implying that
$$u(t)\geq u(0)e^{-2\epsilon_1t}+\frac{\epsilon_3}{\epsilon_1}\left(1-e^{-2\epsilon_1t}\right).$$ 
Hence, $\chi(t)$ is bounded from above as described in the estimate given in \eqref{est1}, from which
 we infer the global existence in time of solutions, that is, $T_{\mathrm{max}}=\infty$ and the limit stated in \eqref{est1}.\\
2. Setting in equation \eqref{feq1}, $\epsilon_1=-\tilde{\epsilon}_1<0$, and working exactly as in case 1, we derive the  estimate of decay given in \eqref{est2}.\\
3. This time we will work with the usual norm of $l^p$-spaces. We set
$$P(t)=h\sum_{n=0}^{N-1}|u_n(t)|^2=||u(t)||_{l^2}^2.$$
and the counterpart of \eqref{feq1}, is
\begin{eqnarray}
\label{feq5}
\dot{P}=2\epsilon_1 P-2\epsilon_2h\sum_{n=0}^{N-1}|u_{n+1}-u_{n}|^2+2\tilde{\epsilon}_3h\sum_{n=0}^{N-1}|u_n|^4.
\end{eqnarray}
The second term of the right-hand side of \eqref{feq5} is estimated from below by the inequality
$$-\sum_{n=0}^{N-1}|u_{n+1}-u_{n}|^2\geq -4\sum_{n=0}^{N-1}|u_{n}|^2,$$
and its third term by the left-hand side of \eqref{equi}, applied again for $q=4$ and $p=2$. Then we have the differential inequality for $P$,
\begin{eqnarray}
\label{feq6}
\dot{P}\geq 2(\epsilon_1-4\epsilon_2)P+2\tilde{\epsilon}_3P^2.
\end{eqnarray}
The Bernoulli inequality \eqref{feq6} can be integrated again with the change of variable $\psi=P^{-1}$, providing the estimate for $P$
\begin{eqnarray*}
\label{feq7}
P(t)\geq \left[-\frac{\tilde{\epsilon}_3}{\epsilon_1-4\epsilon_2}+\left(\frac{1}{P(0)}+\frac{\tilde{\epsilon}_3}{\epsilon_1-4\epsilon_2}\right)e^{-2(\epsilon_1-4\epsilon_2)t}\right]^{-1}=B(t),
\end{eqnarray*}
under the requirement that $B(t)\geq 0$ is non-negative. This condition is satisfied only when,
$$t<\frac{1}{2(\epsilon_1-4\epsilon_2)}\ln\left[1+\frac{\epsilon_1-4\epsilon_2}{\tilde{\epsilon}_3 P(0)}\right]:=T^{*},$$
and $T^*$ is finite, if $\epsilon_1>\epsilon_1^{\mathrm{crit}}$. In this case, solutions exist in the finite time-interval $(0, T_{\mathrm{max}})$ with $T_{\mathrm{max}}\leq T^*$, since
$$\lim_{t\rightarrow T^*}B(t)=\infty,$$
implies the blow-up for $P$. In a similar way, we prove the case $\epsilon_1=4\epsilon_2$, using again the inequality \eqref{feq7}.\ \  $\square$

Note that the critical value $\epsilon_1^{\mathrm{crit}}$ defined in \eqref{estim3} may separate finite time collapse from global existence which is expectable when $\epsilon_1\leq \epsilon_1^{\mathrm{crit}}$ since $T_{\max}\rightarrow +\infty$ as $\epsilon_1\rightarrow \epsilon_1^{\mathrm{crit}}$. Note that collapse may be observed not only in the linear gain/nonlinear gain regime, but also in the linear loss/nonlinear gain regime when $\epsilon_1<0$ and \eqref{estim3} is satisfied.

When $\epsilon_2$ sufficiently small, similar estimates can be proved as in the scenarios 1-3 of Theorem \ref{FAP} which possess essentially a similar functional form with modifications to the exponents governing convergence or the uniform bounds. However,  the dynamics is still governed by the linear and nonlinear gain/loss terms. For example, the estimates of the blow-up time \eqref{estim3}, can be modified to
\begin{eqnarray}
\label{enes1}
T_{\mathrm{max}}&\leq& \frac{1}{2\epsilon_1}\ln\left[1+\frac{\epsilon_1}{\tilde{\epsilon}_3 ||u^0||^2}\right]:=T^*,\;\;\epsilon_1>-\tilde{\epsilon}_3||u^0||^2:=\epsilon_1^{\mathrm{crit}},
\\
\label{enes2}
T_{\mathrm{max}}&\leq& \frac{1}{\tilde{\epsilon}_3||u^0||^2},\;\;\;\epsilon_1=0.
\end{eqnarray}
The functional form of the estimates \eqref{enes1}-\eqref{enes2} remains valid when  $||\cdot||$ is either the standard $l^2$-norm or the averaged norm $\sqrt{\chi(\cdot)}$. With regard to the main result of the closeness Theorem \ref{Theorem:scaledclosenessAL}, the scenarios of Theorem \ref{FAP} are particularly useful in providing estimates of the interval $[0,T_f]$ for which the dissipative solitonic structures may survive when $T_f<T_{\mathrm{max}}$; in particular, such information may have a crucial physical significance in the collapse regime. Of the same usefulness for the dissipative localised structures can be the explicit decay rates in the decay regime, or the spatially averaged energy estimates in the competitive linear gain and nonlinear loss regime. We should stress that the above scenarios describe globally the asymptotic behaviour of the system as they hold for all initial conditions.
\begin{figure}[tbp]
	\begin{tabular}{cc}
		\includegraphics[width=9cm]{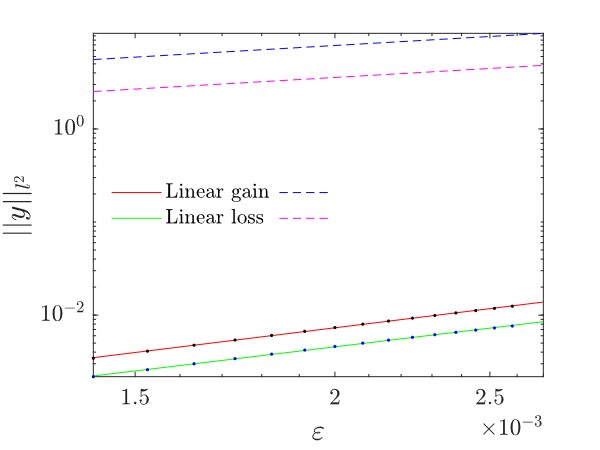} &
		\includegraphics[width=9cm]{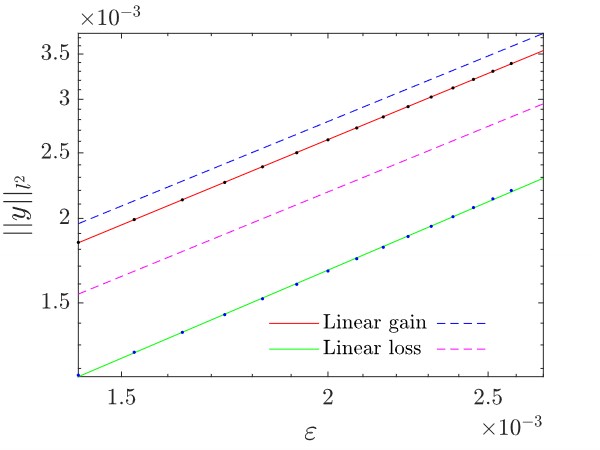} \\
	\end{tabular}%
	\caption{Logarithmic scaled plots of the variation of the distance function $||y(t)||_{l^2}=||u(t)-\psi(t)||_{l^2}$, between the solutions of the DGL \eqref{eq:scaledGLEN} and the AL lattice \eqref{eq:AL} as function of $\epsilon$, for fixed $T_f=500$. Left panel: The case of the dark solitons ($\varepsilon=\sqrt{\sum_n\left(\bigg| |\hat{\phi}_n(0)|^2-|u_n(0)|^2\bigg|\right)/N}$). Right panel: The case of bright solitons ($\varepsilon=\sqrt{\sum_n|u_n(0)|^2/N}$). Details are given in the text.}
	\label{figErrors}
\end{figure}
\paragraph{Dynamics of a wave background: Potential sharpness of the estimates of Theorem \ref{FAP}.}

The estimates proved in Theorem \ref{FAP}, may posses "sharpness properties", as we will justify by the following analysis examining the long-time behaviour of a wave background (in the form of a discrete plane wave), which may support on its top, localised solutions as dark-solitons.  The evolution and stability of a wave background may have important effects on the dynamics of the supported solitonic structures in discrete and continuous set-ups \cite{NGKV,SB0,SB00,SB1,SB2}.

For the DGL equation \eqref{eq:scaledGLEN}, to find the equation for the evolution of the background, we assume the ansatz of solutions
\begin{eqnarray}
	\label{ans1}
u_n = \Phi(t)e^{ i\tilde{k}x_n},\;\;\tilde{k}=\frac{\pi k}{L},\;\;k\in\mathbb{N}.	
\end{eqnarray}
Then inserting the ansatz \eqref{ans1} into \eqref{eq:scaledGLEN}, we  see that $\Phi(t)$ satisfies the complex ODE,
\begin{eqnarray}
	\label{ans2}
i \frac{d\Phi}{dt}=i\epsilon_1\Phi+\Lambda\Phi-(i\epsilon_3-\kappa)|\Phi|^2\Phi,
\end{eqnarray}
with $\Lambda=4(i\epsilon_2-1)\sin^2\left(\frac{\tilde{k}}{2}\right)$.
The second term on the right-hand side of the ODE \eqref{ans2} can be absorbed when using the phase factor
\begin{eqnarray*}
\Phi(t)=e^{i\Lambda t}b(t),	
\end{eqnarray*}
which, if inserted in \eqref{ans2}, implies that $b(t)$ satisfies the equation
\begin{eqnarray}
	\label{ans3}
i\frac{db}{dt}-\kappa|b|^2b=i\epsilon_1b-i\epsilon_3|b|^2b.	
\end{eqnarray}
Still in \eqref{ans3}, the second term on its left-hand side can be absorbed, when using the polar expression for $b$
\begin{eqnarray*}
b(t)=\hat{\phi}(t)e^{\i\theta(t)},\;\;\hat{\phi}:\mathbb{R}\rightarrow\mathbb{R},\;\;\frac{d\theta}{dt}=\hat{\phi}^2,	
\end{eqnarray*}	
deriving this way, the initial value problem for the scalar ODE,
\begin{eqnarray}
	\label{ans4}
	\frac{d\hat{\phi}}{dt}&=&\epsilon_1\hat{\phi}-\epsilon_3\hat{\phi}^3,\\
\label{ans5}
\hat{\phi}(0)&=&\hat{\phi}_0,
\end{eqnarray}
where the initial condition $\hat{\phi}^2(0)=\hat{\phi}^2_0$ defines the initial amplitude of the background (e.g., as induced by the background of the dark soliton initial condition).
The explicit solution of the problem \eqref{ans4}-\eqref{ans5} is
\begin{eqnarray}
	\label{ans6}
\hat{\phi}^2(t)=\frac{\epsilon_1\hat{\phi}_0^2e^{2\epsilon_1t}}{\epsilon_1-\epsilon_3\hat{\phi}_0^2+\epsilon_3\hat{\phi}_0^2e^{2\epsilon_1t}}.
\end{eqnarray}
From the explicit solution \eqref{ans6}, we may deduce the following scenarios for the asymptotic behaviour of the background, which are in compliance with those described in Theorem \ref{FAP}:
\begin{enumerate}
\item
Linear gain $\epsilon_1>0$ and nonlinear loss $\epsilon_3>0$. In this case, the background converges to a finite amplitude,
\begin{eqnarray}
\label{sen1}
\lim_{t\rightarrow\infty}\hat{\phi}^2(t)=\frac{\epsilon_1}{\epsilon_3}.
\end{eqnarray}
The limit \eqref{sen1} is exactly the upper bound for the spatially averaged energy given in the superior-limit of \eqref{est1} stated in Theorem \ref{FAP} and the solution \eqref{ans6} describes the convergence rate of the upper-bound given in \eqref{est1}.
\item
Linear loss $\epsilon_1=-\tilde{\epsilon}_1<0$ and nonlinear loss $\epsilon_3>0$. In this case the background decays, $$\lim_{t\rightarrow\infty}\hat{\phi}^2(t)=0,$$
and the solution \eqref{ans6} describes exactly the  rate of decay given in \eqref{est2} of Theorem \ref{FAP}.
\item Nonlinear gain $\epsilon_3=-\tilde{\epsilon}_3<0$. In this case the background collapses (blows-up) when the linear gain/loss strength satisfies
\begin{eqnarray}
\label{ans7}
\epsilon_1>\epsilon_1^{\mathrm{crit}}:=-\tilde{\epsilon}_3\hat{\phi}^2_0.	
\end{eqnarray}
in the finite time
\begin{equation}
	\label{ans8}
	T_{\max}=\frac{1}{2\epsilon_1}\ln\left[1+\frac{\epsilon_1}{\tilde{\epsilon}_3 \hat{\phi}^2_0}\right].
\end{equation}
In the case where $\epsilon_1=0$, the background blows-up in finite time
\begin{equation}
	\label{ans9}
	T_{\max}=\frac{1}{\tilde{\epsilon}_3\hat{\phi}_0^2}.
\end{equation}
\end{enumerate}
The condition \eqref{ans7} and the estimates of the blow-up time $T_{\max}$  \eqref{ans8}-\eqref{ans9} are exactly the ones given in \eqref{enes1}-\eqref{enes2}, for the case $\epsilon_2\leq 0$, since for the ansatz of plane-wave solutions, the linear coupling terms are absorbed as described in the derivation of the ODE initial value problem \eqref{ans3}-\eqref{ans4}.

The scenarios of Theorem \ref{FAP} for the global asymptotic behaviour of solutions of the finite lattice approximations and the above analysis of the dynamics of a wave background will be important in explaining the dynamics observed in the numerical simulations, that will be reported in the next subsection.
%
\begin{figure}[tbp]
	\begin{tabular}{cc}
		\includegraphics[width=9cm]{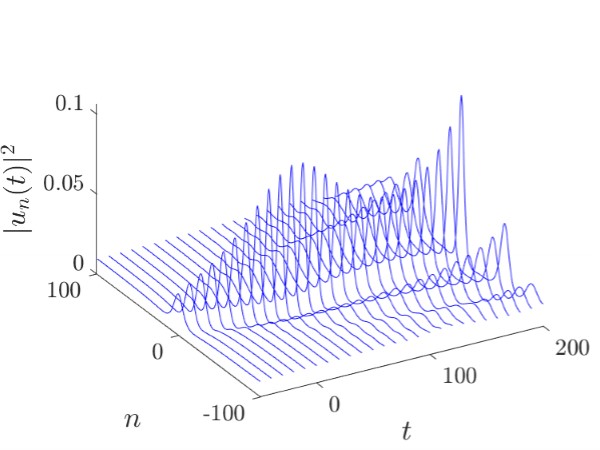} &
		\includegraphics[width=9cm]{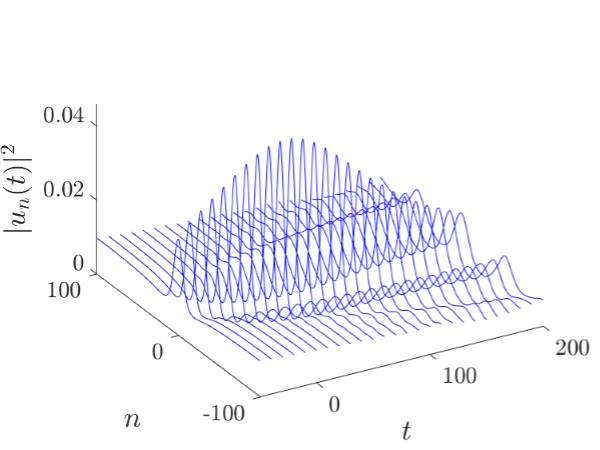} \\
	\end{tabular}%
	\caption{Focusing case $\kappa<0$. Spatiotemporal evolution in the DGL equation \eqref{eq:scaledGLEN} with $\epsilon_2=0.001$ of the initial condition $u_n(0)=\psi_n^r(-50)$ defined by the the analytical Peregrine soliton \eqref{prw_exact} of the AL lattice \eqref{eq:AL},  for $q=0.1$. Left panel: linear gain $\epsilon_1=0.001$ and nonlinear loss $\epsilon_3=0.001$. Right panel: linear loss $\epsilon_1=-0.001$ and nonlinear loss $\epsilon_3=0.001$. Details in the text-see Section $\mathrm{B.c.1}$ and $\mathrm{B.c.2}$.) }
	\label{fig:evolution_Peregrine}
\end{figure}
%
\subsection{Numerical results}
In the light of the analysis provided in the previous subsection, we present in this subsection the numerical results. The numerical study concerns three types of initial conditions provided by the analytical solutions of the AL-lattice \eqref{eq:AL}: dark solitons, bright solitons  and discrete Peregrine solitons (PS). In all cases, if not stated otherwise, we consider in the DGL-lattice \eqref{eq:scaledGLEN} parameter values $|\kappa|= 1$, $\epsilon_2>0$, with $|\epsilon_i|=0.001$, $i=1,2,3$.
\paragraph{Defocusing case $\kappa>0$: Dark Solitons.}
\label{DSa}
In this section we present the results concerning the dynamics ensuing from dark soliton initial conditions provided by the analytical solution of the AL-lattice \eqref{eq:AL},
\begin{equation}
\begin{split}
\psi^d_n(t)&=A\mathrm{tanh}\left[\beta(n-ct)\right]\exp(-i(\omega t-\alpha n)),\;\;A=\tanh\beta,\\
\;\;\omega&=\cos \alpha \sech \beta,\\
c&=-\beta^{-1}\sin \alpha \tanh \beta,\label{eq:dark-soliton}
\end{split}
\end{equation}
with $\alpha \in [-\pi,\pi]$ and $\beta \in [0,\infty)$.
\begin{enumerate}
	\item {\em Linear gain $\epsilon_1>0$ against nonlinear loss $\epsilon_3>0$}.
	\label{DSa1}	
We use as initial condition $u_n(0)=\psi_n^d(0)$ for $\beta=0.01$ and $\alpha=\pi/10$. The top left panel of Figure \ref{fig:evolution_dark} shows the time evolution of the dark soliton density $|u_n|^2$, and the top-right panel, the  normalised density $|u_n|^2/\hat{\phi}^2(t)$, where $\hat{\phi}^2(t)$ is given by \eqref{ans6}. The top panels illustrate the evolution for $t\in [0, 1500]$ and $n\in [-4000, 4000]$. We observe a robust evolution of the dissipative dark soliton as guaranteed by the closeness result of Theorem \ref{Theorem:scaledclosenessAL} on the deformed background by the presence of the gain/loss effects. In particular, the soliton's evolution is only modified by the dynamics of the background, as it is illustrated by the evolution of the normalised density. Both figures corroborate the accuracy of the analytical predictions described by the case 1 of Theorem \ref{FAP} and case 1 for the evolution of the wave background: the dissipative soliton profile is remarkably similar to the analytical dark soliton \eqref{eq:dark-soliton} of the AL-lattice and traces the path guided by the evolution of the background dictated by the ODE-solution \eqref{ans6}. This fact is even more evidentially depicted in the bottom left panel of Figure \ref{fig:evolution_dark}. It is  clearly discernable  that the evolution of the background for the dark soliton initial condition, depicted by the continuous (blue) curve,  follows exactly the dynamics described by the estimates given in \eqref{est1}, reaching an asymptotic state for its amplitude as given in \eqref{est1}. For the chosen parameter values this asymptotic state should be $\leq$ $\epsilon_1/\epsilon_3=1$. We stress that the behaviour described by the estimates \eqref{est1} holds for all initial conditions. Thus, for the dark soliton initial data described above, one should expect the asymptotic limit to be $<\epsilon_1/\epsilon_3=1$. The strict equality for the limit $=\epsilon_1/\epsilon_3=1$ should be expected  for plane waves of the form \eqref{ans1} or spatially homogeneous initial data, where the dynamics are described by the exact ODE initial-value problem \eqref{ans4}-\eqref{ans5}. This fact is illustrated by the numerical result for the evolution of a plane wave  with the same amplitude as of the background of the dark-soliton initial condition, depicted by the dashed (red) curve. The dynamics is exactly the one given by the solution of the ODE initial value problem, given in  \eqref{ans6} and its asymptotic limit goes to $\epsilon_1/\epsilon_3=1$ as expected from \eqref{sen1}.

Concerning the behaviour of the dark soliton on the deformed background, the former evolves such that it eventually reaches  the asymptotic state of the latter, which is the global attractor of the system. Its destabilisation occurs after $t\approx 2800$, as seen in the bottom right panel, displaying the relevant evolution of the normalised density for $t\in [2000, 3500]$.	
 \item {\em Linear loss $\epsilon_1<0$ against nonlinear loss $\epsilon_3>0$}.
 This case corresponds to the case 2 of Theorem \ref{FAP} for the decay of solutions and case 2 for the decaying dynamics of the wave background. The decay of the dissipative soliton is illustrated in the left panel of Figure \ref{fig:evolution_dark1}, while the right panel shows the evolution of the normalised density $|u_n|^2/\hat{\phi}^2(t)$, yet justifying that the AL-lattice dark soliton persists in the DGL lattice ``modulo'' its modification by the decaying background.
 \item {\em Blow-up regime: linear gain $\epsilon_1>0$ and nonlinear gain $\epsilon_3<0$}. This case corresponds  to the case 3 of Theorem \ref{FAP} for the blow-up  of solutions and case 3 for the blow-up dynamics of the wave background. Since the parameters $\epsilon_i$ are small, we test the simplified version of the analytical estimates for the blow-up times \eqref{enes1} against the numerical blow-up times by fixing $\epsilon_2=0.001$, $\epsilon_3=-0.001$, varying $\epsilon_1=\mathcal{O}(10^{-3})$. Due to the form of the inequality  \eqref{feq6}, the comparison principle of ODE's \cite{Zei}, applied to the ODE problem \eqref{ans4}-\eqref{ans5} suggests that the blow-up time \eqref{ans9} for plane waves should serve as a lower-bound for the actual blow-up time, while \eqref{enes1} gives an upper bound as proved in Theorem \ref{FAP}. The validity of the estimates is illustrated in the left panel of Figure \ref{fig:blow-up}. The dashed-dotted (green curve) shows the upper bound $T^{*}$ for the blow-up time \eqref{enes1} as a function of  $\epsilon_1$; in the formula \eqref{enes1} we use the averaged $l^2$-norm of the initial data. The solid (blue curve) corresponds to the numerical blow-up time for the dark-soliton initial conditions. The dashed (red curve) plots the corresponding analytical blow-up time curve  $T_{\mathrm{max}}$ \eqref{ans8} for plane waves. In order to evaluate the numerical blow-up time, we have used -as in the whole simulations of the paper- a Dormand-Prince algorithm with a termination event triggered when the $l^2$-norm of the solution is higher than $10^8$. Using higher blow-up thresholds only leads to drastically increase the computation time without noticeable changes in the accuracy of the blow-up time.

 We stress again the fact that the analytical estimate \eqref{enes1} is generic for all initial data and its proximity to the actual blow-up time may depend on the specific initial data. However, such investigations are beyond the scope of the present work. On the other hand, we observe that the blow-up times for plane waves \eqref{ans9} is closer to the numerical ones. This is expectable, since the dark soliton initial condition resembles a density dip on a constant background, and thus, its blow-up dynamics should be closer to the blow-up dynamics of a plane wave. The evolution of the dark soliton initial condition towards blow-up for $t\in [0,2000]$ is depicted in the left panel, while its normalised density is the same as in bottom left-panel of Figure \ref{fig:evolution_dark}.
 \end{enumerate}
Another interesting feature relevant to the closeness result of Theorem \ref{Theorem:scaledclosenessAL} is the comparison of the evolution of the center of mass of the AL-dark soliton, $X_\mathrm{CM}=X_0+ct$, with $c=(2/\beta)\sin\alpha\tanh\beta$, and the center-of-mass of the dissipative soliton defined as
\begin{equation}
X_\mathrm{CM}=\frac{\sum_{n=n_0-\delta}^{n_0+\delta} n (1-|u_n|^2/|\hat\phi|^2)}{\sum_{n=n_0-\delta}^{n_0+\delta} (1-|u_n|^2/|\hat\phi|^2)},
\end{equation}
with $n_0$ being the location of the minimum density $|u_n|^2$ of the dark soliton, and $\delta$ accounts for the size of the soliton core ($\delta\sim100$). The results of the comparison are shown in Figure \ref{fig:CM_dark}.  The left panel corresponds to the dynamics in the linear gain/nonlinear loss regime presented in Figure \ref{fig:evolution_dark} and the right panel belongs to the dynamics in the linear loss/nonlinear loss regime presented in Figure \ref{fig:evolution_dark1}. We observe that in both cases, the paths are almost indistinguishable for significant time intervals. A divergence starts at an earlier time in the loss/loss regime than in the gain/loss regime. This is 
due to the decay in the  loss/loss regime of the background supporting the dark soliton.
\paragraph{Focusing case $\kappa<0$: Bright Solitons.}
\label{DSa}
In this section we present the results concerning the dynamics of dark soliton initial conditions provided by the analytical solution of the AL-lattice \eqref{eq:AL},
\begin{equation}
\begin{split}
\psi^s_n(t)&=A\mathrm{sech}\left[\beta(n-ct)\right]\exp(-i(\omega t-\alpha n)),\;\;A=\sinh\beta,\\
\;\;\omega&=-2\cos \alpha \cosh \beta,\\
c&=2\beta^{-1}\sin \alpha \sinh \beta,\label{eq:one-soliton}
\end{split}
\end{equation}
with $\alpha \in [-\pi,\pi]$ and $\beta \in [0,\infty)$.
\begin{enumerate}
	\item {\em Linear gain $\epsilon_1>0$ against nonlinear loss $\epsilon_3>0$}.
We use as initial condition $u_n(0)=\psi_n^s(0)$ for $\beta=0.01$ and $\alpha=\pi/10$. The top left panel of Figure \ref{fig:evolution_bright} shows the time evolution of the bright soliton density $|u_n|^2$, and the top right panel, the  normalised density $|u_n|^2/\hat{\phi}^2(t)$; both panels illustrate the evolution for $t\in [0, 1500]$ and $n\in [-4000, 4000]$. We observe again a robust evolution of the dissipative bright soliton as guaranteed by the closeness result of Theorem \ref{Theorem:scaledclosenessAL}. Its evolution  is only influenced by the dynamics of the background as in the dark soliton case; the dynamics of the latter is in full agreement  with the one described by the corresponding  ODE-solution \eqref{ans6} as shown in the bottom panels of Figure \ref{fig:evolution_bright}, where the background amplitude converges to the asymptotic state $\epsilon_1/\epsilon_3=1$.
	\item {\em Linear loss $\epsilon_1<0$ against nonlinear loss $\epsilon_3>0$}.
	\label{DSa2}
 The decay of the dissipative  bright soliton is depicted in the left panel of Figure \ref{fig:evolution_bright1}, while the right panel shows the evolution of the normalised density $|u_n|^2/\hat{\phi}^2(t)$. We observe again that the AL-bright soliton persists in the DGL lattice ``modulo'' its modification by the decaying background.
  \item {\em Blow-up regime: linear gain $\epsilon_1>0$ and nonlinear gain $\epsilon_3<0$}. We performed a study for the blow-up dynamics of the bright solitons, similar to the one of the dark solitons. The parameters $\epsilon_i$ are the same as in the study for the dark solitons. The only curve which is absent is the red (dashed) curve $T_{\mathrm{max}}$ \eqref{ans8} for plane waves, since it is irrelevant to the case of vanishing boundary conditions. The validity of the analytical upper bound \eqref{enes1} in regard to the numerical blow-up time is pictured in the left panel of Figure \ref{fig:blow-up2}. In order to evaluate the numerical blow-up time, we used the same numerical technique as for dark solitons.  We observe that the analytical upper bound is not as close to  the numerical blow-up time, as in the case of the dark solitons. We also observe by a comparison of the numerical blow-up times, that the bright solitons blow-up at earlier times than the dark ones. These effects suggest that the spatial and localisation nature of the initial conditions may affect drastically the structure of the blow-up scenario and its associated time.

\end{enumerate}

The results  comparing the temporal evolution of  the  center of masses of the dissipative soliton and the bright AL-soliton are illustrated in Figure \ref{fig:CM_bright}.  Again we  observe that in both cases the paths are almost indistinguishable for significantly long  time intervals.  Note that in the loss/loss regime we do not observe the divergence of paths, as the bright soliton, vanishing as $|n|\rightarrow\infty$, evolves on a "zero" amplitude background which is not affected by the loss/loss dynamics.
\paragraph{Accuracy of the closeness estimates of Theorem \ref{Theorem:scaledclosenessAL}.}
Theorem \ref{Theorem:scaledclosenessAL} establishes the convergence of solutions of the DGL equation \eqref{eq:scaledGLEN} to the solutions of the AL-lattice \eqref{eq:AL}, at a rate of $\mathcal{O}({\varepsilon})$, when $\epsilon\leq\varepsilon\rightarrow 0$, and the distance between their initial conditions is $||u^0-\psi^0||_{l^2}=\mathcal{O}({\varepsilon})$.

Figure \ref{figErrors}, depicts logarithmic scaled plots of the variation of the distances  $||y||_{l^2}=||u(t)-\psi(t)||_{l^2}$  as functions of $\varepsilon$ for fixed $T_f=500$. The left (right) panel illustrates the results of the case of dark (bright) solitons.  The dashed lines in both panels, correspond to lines of the analytical estimates of Theorem \ref{Theorem:scaledclosenessAL} of the form  $||y||_{l^2}$ versus $C\varepsilon$. For the case of the dark solitons, we have $C=3.9\times10^3$ in the linear gain/nonlinear loss regime, corresponding to the top dashed (blue) line. In the case of the linear loss/nonlinear loss regime, we have $C=1.8\times10^3$, which corresponds to the second dashed (purple) line. The dots on the solid lines correspond to the numerically detected rates of the variations of the distance functions fitted to the lines of the form $||y||_{l^2}$ versus $C\epsilon^a$,  for the above given values of the constant $C$.  In the case of the linear gain/nonlinear loss we found that $a=2.12$, as illustrated by the dots on the first solid (red) line. In the case of linear loss/nonlinear loss, we found that $a=2.07$, as depicted by the dots on the second solid (green) line. The numerical results illustrate that the analytical estimates are not only fulfilled, but also that the numerical variation of the distance functions is of significantly lower rate, namely of order $\sim\epsilon^2$.

In the case of the bright solitons, for the analytical estimates we found for the  dashed  curves, that $C=1.0336$ in the case of linear gain/nonlinear loss regimes and $C=1.0945$ for the linear loss/nonlinear loss regimes. For the numerically determined rates, we found $a=1.01$ in the case of the linear gain/nonlinear loss regime and $a=1.04$ for the linear loss/nonlinear loss regime, very close to the analytical predictions for the convergence estimates of order $\sim\varepsilon$.

As a conclusion, while the analytical closeness estimates are satisfied, the numerical studies indicate that the actual order of closeness may depend  on the specific nature of the initial conditions, e.g., the type of their localisation. In this context, whether an improvement of the theoretical estimates is achievable, is an interesting problem to be explored in future investigations.

\paragraph{Focusing case $\kappa<0$. Discrete Peregrine Solitons.}
Finally,  we present the results concerning the dynamics ensuing from discrete Peregrine initial conditions provided by the  analytical  Peregrine soliton solution of the AL-lattice \eqref{eq:AL},
\begin{equation}
\psi_{n}^r(t)=q\left[1-\frac{4(1+q^2)(1+4iq^2t)}{1+4n^2q^2+16q^4t^2(1+q^2)}\right]e^{2iq^2t},
\label{prw_exact}
\end{equation}
where the parameter $q$ fixes a  background amplitude.
\begin{enumerate}
	\item {\em Linear gain $\epsilon_1>0$ against nonlinear loss $\epsilon_3>0$}.
	We use as initial condition $u_n(0)=\psi_n^r(-50)$ with a background amplitude $q=0.1$. We observe in the left panel of Figure \ref{fig:evolution_Peregrine}, that  the profile of the analytical Peregrine soliton is preserved up to $t\approx 50$. As it can be expected, the structures are not temporarily localised: a rise of amplitude follows since the dynamics is  chiefly controlled  by the evolution of the background, governed by the solution of the ODE \eqref{ans6}.
	\item {\em Linear loss $\epsilon_1<0$ against nonlinear loss $\epsilon_3>0$}.
The loss/loss regime seems to be more physically relevant to the construction of discrete spatio-temporally localised waveforms reminiscent of the Peregrine soliton as the solution in this regime eventually decays. Hence, we may expect that a spatio-temporal waveform may survive prior to the eventual decay dynamics. This very scenario is depicted in the right panel of \ref{fig:evolution_Peregrine}. In fact, prior to  the eventual decay of the solution, the dynamics ensuing from the initial condition  $u_n(0)=\psi_n^r(0)$ exhibits a spatio-temporal localised pattern, akin to that of the analytical AL Peregrine soliton.
\end{enumerate}
\section{Conclusions}
Introducing a dynamical transitivity argument, which combines the notions of ``inviscid limit''  and ``continuous dependence on their initial data" between dissipative and Hamiltonian integrable and non-integrable nonlinear lattices \cite{DNJ2022}, we have proved the persistence of localised structures for the discrete Ginzburg-Landau equation, for small values of its dissipation or gain strengths. The persisting waveforms are close to the analytical solutions of the Hamiltonian integrable Ablowitz-Ladik lattice, with their distance measured in the suitable metrics induced by the discrete ambient space, when their initial data are also close.
The numerical simulations confirm the main closeness theoretical result, namely that the localised structures in the form of bright and dark solitary waves on the Ginzburg-Landau lattice  share for significant times major characteristics, such as the functional form and  velocity, with their conservative counterparts. Moreover, in full agreement  with a systematic analysis for the finite lattice approximations (relevant to the numerical studies), the numerical findings illustrate that the global asymptotic behaviour of the  dissipative solitary waves is mainly controlled by the dissipation or loss effects which are present in the discrete Ginzburg-Landau system. We remark that these results have an impact in determining the stability of even the exact solutions of the discrete Ginzburg-Landau system \cite{EX1,EXA1,EX2}, for which, if their small perturbations will be used as initial conditions in the system, should exhibit the global asymptotic behaviour identified herein.  However, an important feature of the dynamics is that the Ablowitz-Ladik solitons persist in the dissipative system  "modulo" the growth/decay rates (which  are analytically quantified)  for the dynamics of the wave background determined by the dissipative or energy gain effects. Furthermore, we have studied the existence and persistence of spatiotemporally localised waveforms possessing the characteristics of the discrete Peregrine solitons in discrete DGL systems. These results may suggest other applications such as the persistence of the solitary waves without "deformation" when in the system additional suitable terms \cite{EX2} are included. Another direction will investigate potential extensions of the closeness argument to continuous systems. Investigations in this direction are in progress and relevant results will be reported elsewhere \cite{DNJI2022}.
\section*{Acknowledgment}
\vspace{-0.3cm}
We would like to thank the referees for valuable comments and suggestions. JCM acknowledges support from EU (FEDER program2014-2020) through both Consejer\'{\i}a de Econom\'{\i}a, Conocimiento, Empresas y Universidad de la Junta de Andaluc\'{\i}a (under the projects P18-RT-3480 and US-1380977), and MCIN/AEI/10.13039/501100011033 (under the projects PID2019-110430GB-C21 and PID2020-112620GB-I00).
\vspace{2cm}\\
\textbf{Authors Declarations}\\
The authors have no conflicts to disclose.\\
\\
\textbf{Authors Contributions Statement}\\
All authors contributed equally to the study conception, design and writing of the manuscript. Material preparation, data collection and analysis were performed equally by all authors.  All authors read and approved the final manuscript.

\end{document}